\documentstyle[emulateapj]{article}

\def\Ha{H$\alpha$}
\def\ha{\rm H\alpha}
\def\Dt{\spose{\raise 1.5ex\hbox{\hskip3pt$\mathchar"201$}}}    
\def\cm2{{\rm\,cm^{-2}}}
\def\cc{{\rm\,cm^{-3}}}
\newcommand\hi{\ion{H}{1}}
\newcommand\hii{\ion{H}{2}}
\newcommand\oiii{[\ion{O}{3}]}
\newcommand\sii{[\ion{S}{2}]}
\newcommand\nhi{$N$(\ion{H}{1})}
\newcommand\etal{et\thinspace al.~}

\newcommand\msol{\rm\,M_\odot}

\newcommand\kms{{\rm\,km\,s^{-1}}}

\def\spose#1{\hbox to 0pt{#1\hss}}
\def\simpropto{\mathrel{\spose{\raise 3pt\hbox{$\propto$}}
     \lower 3.0pt\hbox{$\sim$}}}

\received{}
\accepted{}

\slugcomment{Accepted to the Astronomical Journal 2001-October-09}

\lefthead{M. S. Oey {\etal.}}
\righthead{\hi\ imaging of superbubbles}

\begin{document}

\title{	
The \hi\ Environment of Three Superbubbles in the Large Magellanic Cloud
}

\author{M. S. Oey}
\affil{Lowell Observatory, 1400 W. Mars Hill Rd., Flagstaff, AZ   86001,
	USA; oey@lowell.edu}
\author{B. Groves}
\affil{Research School of Astronomy and Astrophysics, Australian
	National University, Private Bag, Weston Creek P.O., 
	ACT 2611, Australia; bgroves@mso.anu.edu.au}
\author{L. Staveley-Smith}
\affil{Australia Telescope National Facility\altaffilmark{2}, CSIRO,
	P.O. Box 76, Epping, NSW 1710, Australia; 
	Lister.Staveley-Smith@atnf.csiro.au}
\altaffiltext{2}{The Australia Telescope Compact Array is part of the
Australia Telescope, which is funded by the Commonwealth of Australia
for operation as a National Facility managed by CSIRO.}
\author{R. C. Smith}
\affil{Cerro Tololo Inter-American Observatory\altaffilmark{3},
	Casilla 603, La Serena, Chile; csmith@noao.edu}
\altaffiltext{3}{National Optical Astronomy Observatories, operated by
the Association of Universities for Research in Astronomy, Inc., under
cooperative agreement with the National Science Foundation, U.S.A.}

\begin{abstract}

The ambient interstellar environment of wind- and supernova-driven
superbubbles strongly affects their evolution, but its properties are
rarely well-determined.  We have therefore obtained \hi\ aperture synthesis
imaging of the environment around three similar, optically-selected
superbubble nebulae in the Large Magellanic Cloud.  The resulting \hi\ maps
show that the ambient gas distribution around these superbubbles
differ to an extreme:  DEM L25 shows no neutral shell component, but
is nestled within an \hi\ hole; DEM L50 shows a massive neutral shell
component, but is otherwise within an \hi\ void; and DEM L301 shows no
correspondence at all between the optical nebula and \hi\ distribution.
There is also poor correspondence between the \hi\ and optical kinematics.
These results strongly caution against inferring properties of the
ambient neutral environment of individual superbubbles without direct
observations.  Finally, all three objects show some evidence of shock
activity. 

\end{abstract}


\section{Introduction}

The evolution of superbubbles created by the supernovae (SNe) and stellar
winds of OB associations is a critical determinant of the structure
and energetics of the interstellar medium (ISM).  The coronal
component of the ISM, or hot ionized medium (HIM) is believed to
originate within such superbubbles and supernova remnants (SNRs; e.g., Cox
\& Smith 1974), hence the evolution of these structures determines
whether and how this hot gas is released into the diffuse HIM.  Models
for the ISM are strongly dependent on whether the HIM is a pervasive,
dominant component, as envisioned by e.g., McKee \& Ostriker (1977),
or whether it plays a lesser role as favored by, e.g., Slavin \& Cox
(1993), returning to a paradigm closer to the two-phase ISM of Field,
Goldsmith, \& Habing (1969).  Superbubble evolution also directly
impacts the relative volumes of enriched, coronal gas in galactic
disks vs. halos (e.g., Shapiro \& Field 1976).  Similarly,
superbubbles and their related filaments are thought to provide much
of the structure for the diffuse ISM, including warm and cold \hi\ and
the warm ionized medium (e.g., Kennicutt {\etal}1995).  
Superbubble activity is therefore fundamental to the ISM phase balance,
structure, and kinematics; and it is also a principal driver of galactic
evolutionary processes.

The standard model for the evolution of these superbubbles is an
adiabatic, pressure-driven shell with continuous mechanical energy
injection (e.g., Pikel'ner 1968; Castor, McCray, \& Weaver 1975).
The mechanical power $L$ is provided by stellar winds and SNe of the
parent OB association.  In this model, an outer shock sweeps up
ambient ISM into a thin cooled shell, while an inner, reverse shock
thermalizes the mechanical energy, thereby generating the hot interior
whose pressure drives the shell growth.  It is generally assumed that
successive SNe can be approximated as a continuous energy injection,
and that all the available SN energy is thermalized (e.g., Mac Low \&
McCray 1988).  The standard, adiabatic model then yields simple
analytic relations governing the shell evolution:
\begin{equation}
R\propto \bigl(L/n\bigr)^{1/5}\ t^{3/5} \quad ,
\end{equation}
\begin{equation}
v = \Dt{R}\propto \bigl(L/n\bigr)^{1/5}\ t^{-2/5} \quad ,
\end{equation}
where $R$ and $v$ are the shell radius and expansion velocity,
respectively, $n$ is the ambient density, and $t$ is elapsed time.

There are several lines of evidence supporting this standard model for
superbubble evolution (see Oey 1999):  a few young, stellar
wind-dominated objects have
well-constrained $L$ from classification of the stellar populations,
and show reasonable consistency with the predicted kinematics (Oey
1996b).  In addition, multiwavelength observations of highly active
star-forming regions like 30 Doradus (Wang 1999) and the starburst
galaxy NGC 5253 (Strickland \& Stevens 1999) show hot gas
morphologically contained within cooler shell structures.  
Higher ionization species like \ion{C}{4} and \ion{Si}{4} have been
detected in lines of sight through all superbubbles with available
observations (Chu {\etal}1995), suggesting interface zones between hot
and cold gas.  In addition, the superbubble size distribution
derived from the standard model is in remarkable agreement with
observed size distributions of \hi\ shells in the Magellanic Clouds
(Oey \& Clarke 1997; Kim {\etal}1999).

However, other evidence suggests that superbubble evolution is often
not nearly as simple as described by the standard model.  The first
problem is that the model overestimates the shell growth rate for
young, wind-dominated superbubble nebulae (Oey
1996b; Garc\'\i a-Segura \& Mac Low 1995; Drissen {\etal}1995).  This
could be due to a systematic understimate in the ambient densities $n$
by up to a factor of $\sim10$, or to a similar overestimate in the input
power $L$ predicted by the assumed stellar mass-loss rates.  Secondly,
Oey (1996b) identified a class of superbubbles that 
exhibit expansion velocities that are several times
higher than predicted, relative to their radii.  Most of these also
show anomalously bright X-ray emission, which
suggests shell acceleration by SNR impacts (Chu \& Mac Low 1990; 
Oey 1996b).  In such cases, not all the available SN energy will be
thermalized to power the shell expansion, thus the effect of discrete
SNe on shell evolution remains unclear.

Clarifying the structure of the ambient medium is clearly essential in
determining the origin of these discrepancies.  Since most estimates
of the ambient $n$ for young nebular shells are based on the \Ha\
electron densities, the presence of a massive \hi\ shell component
could imply a significant underestimate in assumed $n$.  Another
possibility is that the structure of the ambient medium is highly
irregular, providing openings to a network of low-density channels.
In that case, the superbubble could be ruptured in one or more
locations, depressurizing the interior and releasing hot gas into
these channels.  Soft X-ray imaging with {\it Einstein} and {\it
ROSAT} suggests this process in at least one object, DEM L152 (Chu
{\etal}1993; Magnier {\etal}1995) in the Large Magellanic Cloud
(LMC).  These minor blowouts would then cause accelerated,
high-velocity kinematics in excess of the standard model predictions
(e.g., Oey \& Smedley 1998; Mac Low {\etal}1998).  Expansion into
the density gradient perpendicular to galactic disks offers an additional
systematic explanation for anomalously fast shell expansion in
face-on galaxies (Silich \& Franco 1999).

\section{Observations}

\subsection{Optical narrowband imaging}

In order to examine their interstellar environment in more detail, we
therefore obtained \hi\ imaging around three nebular superbubbles in
the LMC.  All three objects are among those found by Oey (1996b) to show 
anomalously high expansion velocities relative to their radii.
The characteristics of these objects are summarized in
Table~\ref{tabsample}:  columns 1 and 2 give their identifications in the
\Ha\ catalogs of Davies, Elliott, \& Meaburn (1976) and Henize (1956),
respectively; and column~3 gives the shell radius $R$ from Oey (1996b),
assuming a 50~kpc distance to the LMC.
All three objects contain OB associations in projection, and their stellar
content was classified by Oey (1996a).  The stars with the earliest
spectral types within each superbubble are listed in column~4.  
Finally, column 5 lists the ratio of predicted to observed \Ha\
luminosity $L_{\ha}({\rm pred})/L_{\ha}({\rm obs})$, determined by Oey
\& Kennicutt (1997) from the observed stellar population.

\begin{deluxetable}{llclc}
\tablecaption{Superbubble sample \label{tabsample}}
\tablewidth{0 pt}
\tablehead{
\colhead{DEM} & \colhead{Henize}   & \colhead{$R$(pc)\ \tablenotemark{a}} &
\colhead{Sp. Type\tablenotemark{b}} & 
\colhead{$L_{\ha}({\rm pred})/L_{\ha}({\rm obs})$\ \tablenotemark{c}}
} 
\startdata
DEM L25  & N185    & 43 & O9 V        & 8.6 \\
DEM L50  & N186 CE & 50 & O6.5 V((f)) & 2.0 \\
DEM L301 & N70     & 53 & O3 If*      & 0.2 \\
\enddata
\tablenotetext{a}{From Oey (1996b).}
\tablenotetext{b}{From Oey (1996a).}
\tablenotetext{c}{From Oey \& Kennicutt (1997).}
\end{deluxetable}

We obtained optical narrowband imaging of these objects in the light
of \Ha, \oiii$\lambda$5007, \sii$\lambda\lambda$6717,6731, and also
red and green continuum.  These
observations were obtained from the Magellanic Clouds Emission-Line
Survey (MCELS; Smith {\etal}1999), which is nearing completion at the
CTIO/U. Michigan 0.9-m Curtis Schmidt Telescope.  The observations for
DEM L301, which are also part of this survey, were published by Oey
{\etal}(2000), and similar narrowband imaging was reported by Skelton
{\etal}(1999).  The filter bandwidths for \Ha, \oiii, and \sii\ are
30\AA, 40\AA, and 50\AA, respectively, with wider continuum filters
(see Smith {\etal}1999); the spatial resolution is about $3-4\arcsec$,
with 2.3\arcsec~px$^{-1}$.  Figure~\ref{figopt} shows color
composite images constructed from the continuum-subtracted optical
narrowband observations.  \Ha\ is shown in red, \oiii\ in green, and
\sii\ in blue.  The nebular gas in all three objects is well-defined,
highly circular, and filamentary.
The MCELS narrowband frames of DEM L25 and DEM L50 were flux-calibrated
with two longslit spectra of positions in DEM L25.  These
were obtained at the CTIO 1.5-m telescope on 2001 August 1 under
photometric conditions.

\subsection{\hi\ and continuum observations}

We used the Australia Telescope Compact Array (ATCA) to construct aperture
synthesis imaging of these regions in the 21-cm \hi\ line.  As seen in
Figure~\ref{figopt}, DEM L25 and L50 are separated by only $\sim
0.5^\circ$, so for these targets we 
mosaicked a single field from 13 pointings.  Our field for DEM L301
comprises 7 pointings.  We obtained the observations in four
configurations:  750A (1996 Nov 27), 750D (1997 Jan 5--7), 1.5C (1997
Sep 3), and 1.5D (1997 Mar 26--30).  Excluding the data from the
baselines involving antenna~6, these configurations yield 40 independent
baselines ranging from 30 to 1470 m.  The 21-cm line observations were
carried out for an LMC heliocentric radial velocity 
$v_{\rm hel}\sim 297\ \kms$.  The bandwidth of 4 MHz was divided into
1024 channels, which were then binned to 512 channels.  These
correspond to a velocity coverage $-33 < v_{\rm hel} < 627\ \kms$,
with resolution 1.65 $\kms$ and sensitivity of 2.2 K.
We simultaneously observed the 20-cm, 1.380 GHz continuum emission
with a bandwidth of 128 MHz and full polarization information.

We observed the primary flux calibrator PKS B1934--638 at the start
and end of each observing run.  For phase and amplitude calibration,
we observed PKS B0454--810 and PKS B0407--658.  To increase the $uv$
coverage in the shortest baselines, we also merged our data with
fields from the LMC ATCA survey of Kim {\etal}(1998) and the Parkes
single-dish survey (Staveley-Smith {\etal}2001).  For DEM L25 and L50,
we included 50 
fields from the ATCA survey, and for DEM L301, we included 37 fields.
The pointing positions of our new data were adopted to coincide with
the central pointings from the survey data.  Altogether, these
observations yield \hi\ imaging at 50\arcsec\ resolution within roughly
$40\arcmin$ of each object.

The calibration and mosaicking of the radio data were carried out with
{\sc Miriad} software.  Our data were transformed to the image plane
using superuniform weighting and a Gaussian taper of 40\arcsec.  We
deconvolved the beam using a maximum entropy method, and this was then
merged with the ATCA data and regridded Parkes data. 
We adopted a flux ratio between the Parkes data and our data of 1.2
for DEM L25 and L50, and 1.0 for DEM L301.

\section{Results}

The optical
emission-line properties of DEM L301 are discussed in detail by Oey
{\etal}(2000), who also present the individual narrowband images of
that object used here.  Similar CCD narrowband imaging of DEM L301 was 
obtained and discussed by Skelton {\etal}(1999).  In addition,  
Oey {\etal}(2000) obtained longslit spectrophotometry and 
compared these with tailored photoionization and shock models,
showing that the observed line ratios appear to
result from a combination of shock and density-bounded
photoionization.  

In contrast, DEM L25 and L50 emit substantially more \Ha\ luminosity
than apparently can be provided by the stars (Table~\ref{tabsample}).
This implies an additional ionization mechanism, most plausibly shock
heating.  Figure~\ref{figopt}$a$ shows a dramatic contrast between DEM
L25 and L50 in \oiii\ emission:  while DEM L25 shows a ratio of
\oiii$\lambda5007$/\Ha\ $\sim 0.8$, the main shell of DEM L50 shows almost no
detection in \oiii.  There is some \oiii\ emission from the SNR
N186~D, which is the small loop at the north end of DEM L50
(see \S 3.2 below), but Lasker (1977) cites the lack of \oiii\
emission in the main shell as evidence that N186~D is physically
separate from the remainder of the system.  
The \oiii\ emission from DEM L25 is surprising in view of the stellar
populations:  DEM L25 has only a single O9~V star, whereas
DEM L50 has one O6.5~V star and several late-type O stars.  We would
have expected \oiii\ emission from both nebulae, since it is even seen
in \hii\ regions dominated by O9 stars (Hunter 1992; Hunter \& Massey
1990).  However, the extremely low ionization parameter in these shell
structures is likely to quench the photoionized \oiii\ emission in
these objects (e.g., Oey {\etal}2000; Garnett \& Kennicutt 1994).  The
\oiii\ from DEM L25 is therefore most likely caused by an additional
heating mechanism, probably shock excitation.

The nebular kinematics of all three superbubbles have been studied
extensively (e.g., Rosado {\etal}1981; Lasker 1981; Dopita
{\etal}1981; Rosado {\etal}1982; Meaburn, McGee, \& Newton 1984;
Rosado {\etal}1990).  All three show highly complex nebular velocity
structure, and do not exhibit simple uniform expansion.  The complex
kinematics were cited as evidence consistent with the suggestion of
internal SNR impacts to explain the anomalously high velocities (Oey
1996b). 


Figure~\ref{fignhi} shows the wide-angle maps of \hi\ column density
\nhi\ around these regions.  The circles indicate the positions of the
superbubbles at their approximate angular size.
Figure~\ref{figopthi} gives a closer look with composite 
images of the optical and \hi\ data.  Red is \Ha, green is \oiii, and
blue is now \hi.
It is apparent that the neutral environment varies dramatically
among the three objects.  DEM L25 and L50 appear to be situated
in a large \hi\ void, roughly 1.4 kpc in diameter, which apparently
corresponds to the supergiant shell LMC~8 (Meaburn 1980).  This
structure is also clearly 
seen in the full \hi\ map of the LMC contructed by Staveley-Smith
{\etal}(2001), which combines data from the Kim {\etal}(1998) ATCA
survey and the Parkes multibeam receiver data.
Within this void, there is a patch of \hi\ emission
corresponding to DEM L50, but no apparent \hi\ associated with DEM
L25.  Figure~\ref{figd25chan} shows \hi\ maps of the region over six velocity
intervals, which are the average over four channels.  The mean
velocity is shown in the upper left of each panel.
The environment of DEM L301 (Figure~\ref{figopthi}$b$) shows
neutral gas throughout the region in projection, but our data again show no
apparent correspondence in the \hi\ distribution with the nebular
superbubble (see below).  Figure~\ref{figd301chan} shows \hi\ maps of
this region averaged over five velocity channels.
Hence, despite the similar appearance of the three
shells in \Ha, their \hi\ environments are different to an extreme.

\subsection{DEM L25}

As mentioned above, Oey \& Kennicutt (1997) found that both DEM~L25
and DEM~L50 show more \Ha\ emission than can be accounted for by the
late-type O stars present within these objects.  Both nebulae might
therefore be expected to be radiation-bounded and associated with
neutral gas.  The environment of DEM L25 is consistent with the
superbubble being nestled within a partial \hi\ hole (Figure~\ref{figd25}),
within a region where typically \nhi $\sim 1.0\times 10^{20}\ \cm2$. 
While some of the emission seen to the NE of DEM L25 probably belongs
to an unassociated velocity component, the individual channel maps do
remain consistent with general \hi\ anticorrelation for the
superbubble (Figure~\ref{figd25chan}).

The \hi\ scale height $h$ of the LMC is dependent on radius, and difficult
to determine because of the galaxy's low inclination angle.  Kim
{\etal}(1998) estimated a characteristic $h\sim 600$ pc, which is
weighted toward outer radii and assumes that the stars and gas are
coupled.  A more classical analysis by Kim {\etal}(1999) suggests
$h\sim 180$ pc, assuming stars and gas are not coupled.  Elmegreen,
Kim, \& Staveley-Smith (2001) note a break in the \hi\ power spectrum
around a scale of 100 pc, which may also indicate a low value of $h$.
Here we adopt a value of $h = 300$ pc as a compromise estimate.
If the value of \nhi\ found in the DEM L25 environment
is characteristic of the ISM in this region, our adopted $h$
implies an ambient volume density of $n(\sc Hi)\sim 0.05\ \cc$.  This low
value is related to the location of DEM L25 within the supergiant
shell environment mentioned above.  

There is no
emission that unambiguously could correspond to a neutral shell of
swept-up material, thus it may be that the entire shell is ionized.
As seen in Figure~\ref{figopt}, DEM~L25 shows the highest excitation
among the three objects, with \oiii$\lambda5007$/\Ha\ $\sim 0.8$, despite
the fact that DEM L25 has the latest-type OB stars of the three nebulae.
The ratio of observed to predicted \Ha\ luminosity is so high,
$L_{\ha}({\rm obs})/L_{\ha}({\rm pred})= 8.6$ (Table~\ref{tabsample}),
that an additional ionization mechanism is unavoidable.  This is
consistent with the high \oiii/\Ha\ ratio, and 
it is therefore less surprising to find no neutral gas
component.

Figures~\ref{figopthi}$a$ and \ref{figd25} show that on the west side
of DEM L25,  the 
superbubble  is interacting with an \hi\ cloud.  This is apparent
both morphologically and from the enhanced optical nebular emission,
indicating a density increase in the optical shell as would be
expected from a collision.  The \sii/\Ha\ ratio is also higher in the
interaction zone, ranging around $\sim 1.1$,
as would be expected from a ram pressure-induced shock.
The mass of the \hi\ cloud is $\sim$1.3$\times10^4\ \msol$, as
measured within a 230\arcsec\ box centered at 04 53 49.4, --70 00 12
(J2000).  Figure~\ref{figd25pvcloud} shows an
E-W position-velocity (PV) slice across the cloud, showing essentially a
single component with a systemic velocity of 240 $\kms$.
The velocity dispersion at the interaction zone is about twice as
large as for the remainder of the cloud, as is qualitatively expected
from a collision. 
The surrounding \hi\ spatial distribution suggests that the shell is
expanding into the cloud, or rather than that the entire superbubble has a
systemic velocity in collision with the cloud.
The standard, wind-driven evolution predicts a shell expansion velocity
$v_s \simeq 15\ \kms$, based on the observed stellar population and shell
size (Oey 1996b).  This value represents a rough lower limit to the
collision velocity, since \Ha\ kinematic observations by Rosado
{\etal}(1982) show much higher velocity structures that may imply
$v_s$ up to 70 $\kms$.  We note that no kinematic features are seen in
the \hi\ observations at these high velocities, again arguing that the
shell itself is fully ionized.

To the N and NE of DEM L25 is denser \hi\ gas
(Figure~\ref{figopthi}$a$), which is a southern extension from a
major star-forming region with high-density gas about 300 pc to
the north (Figures~\ref{fignhi}$a$ and \ref{figd25}).
Figure~\ref{figd25pvne} shows the 
PV diagram for a N-S slice at position 04 54 28.9, --69 56 21.0,  to
the NE of the shell.  Two distinct kinematic components are clearly
apparent around 245 and 270 $\kms$, which undoubtedly correspond to
those seen in the unresolved observations by Meaburn {\etal}(1984).
These two components are continuously present over most of the \hi\
emission in the large mass immediately to the NE of DEM L25 and contrast with
the single component of the cloud to the W.  The existence of two,
continuous, distinct velocity components is puzzling, but is a common
feature that is also seen, for example, in the Small Magellanic Cloud
(Staveley-Smith {\etal}1997).  

\subsection{DEM L50}

DEM L50 shows an \hi\ configuration that is virtually
inverse to that of DEM L25.  DEM L50 clearly shows associated \hi,
plausibly in a neutral shell envelope outside the nebula, while the
surrounding environment has a low density. 
For a single $1.65\ \kms$ channel, our 3$\sigma$ limit is 2.0$\times
10^{19}\ \cm2$, although the effective detection limit depends on the
kinematic range of the structure.  Binning over four channels
($\Delta v= 6.6\ \kms$) as shown in, e.g., Figure~\ref{figd25chan}
gives a nominal detection limit of 4.0$\times 10^{19}\ \cm2$.
This limiting ambient \hi\ column density
corresponds to a volume density of $n(\sc Hi) \lesssim 0.02\ \cc$, an
extremely low value that is again associated with the interior of the
supergiant shell environment.  Figure~\ref{figd50} shows the excellent spatial
correspondence of the \Ha\ (image) and \nhi\ (contours) maps.  The total
integrated emission of 282 Jy $\kms$ within a 610\arcsec\ box implies a total
\hi\ mass of 1.7$\times10^5\msol$ for this material.  For comparison,
the ambient density of 1.4 $\cc$ estimated by Oey (1996b) and 50 pc
radius imply a swept-up mass of 1.6 $\times 10^4\ \msol$.  This
therefore suggests that the ambient density was underestimated
by an order of magnitude from the \Ha\ emission.  Alternatively, it
may that most of the \hi\ originates in a structure exterior to the
superbubble, although this seems unlikely because of the spatial
correlation between the \Ha\ and \hi\ emission in Figure~\ref{figd50}.

Figure~\ref{figd50pv} shows PV diagrams for E-W and N-S
slices across the center of DEM~L50.  Most of the material has a
velocity $\sim 247\ \kms$.  The N-S slice 
(Figure~\ref{figd50pv}$a$) shows a gradient to $\sim 260\ \kms$ near
the edges, suggesting that most of the central shell material
may correspond to a near-side shell expansion velocity of $v_s \sim 13\
\kms$.  Based on the observed stellar population and shell radius, the
standard wind-driven evolution predicts $v_s \sim 20\ \kms$ for the
originally assumed ambient $n = 1.4\ \cc$, and $v_s \sim 16\ \kms$ for $n$
increased by an order of magnitude (Oey 1996b).  Both values are
consistent with a near-side \hi\ expansion.  The \Ha\ velocity
structure is again complex (Rosado {\etal}1990); while the reported
values are broadly consistent with the principal \hi\ component around
$\sim 240\ \kms$, it is difficult to identify the optical kinematic
structures with the \hi.  Rosado {\etal} suggest a possible optical
shell expansion of $v_s\sim 25\pm 9\ \kms$, reasonably consistent with
the tentative near-side \hi\ expansion.  Howver, it is also possible
that the entire superbubble may have a systemic velocity offset from
the ambient ISM, or that the shell kinematics are more complicated.

Figure~\ref{figd50pvsnr} shows a PV diagram for the SNR N186~D, obtained
from a N-S slice at position 05 00 12.3, --70 06 22.  
Here we do clearly see a $\sim 20\ \kms$ line
splitting over a region extending about 4\arcmin.  However, this is
puzzlingly inconsistent with \Ha\ observations suggesting $v_s\sim 90\
\kms$ for the SNR, found in the \Ha\ Fabry-P\'erot data of Laval
{\etal}(1989).  Their data generally appear to show two components
around 160 and 320 $\kms$, and there are no reported velocity structures 
in the range 240 to 270 $\kms$, which are seen in our \hi\
observations.  Our data cube has limiting extremes of 190 and 336
$\kms$, so we cannot adequately probe the same kinematic range as the
optical data, but we detect no structures near these extremes
in our cube, within our effective detection limit of $\sim 4\times
10^{19}\ \cm2$.

\subsection{DEM L301}

The \hi\ environment of DEM L301 is still different in character from
either that of DEM L25 or DEM L50.  As seen in Figure~\ref{figd301},
there is no apparent correspondence at all between the \hi\ emission and
nebular emission, whether as a correlation or anticorrelation.  Nor
can we identify any correspondence in the individual channel maps,
(Figure~\ref{figd301chan}).
Figure~\ref{figd301pv} shows PV diagrams for E-W and N-S slices
across the center of the optical nebula.  There is no kinematic
structure readily identifiable with the superbubble either.
Optically, DEM L301 again shows complex, high-velocity components in
\Ha, with velocity differences up to 100 $\kms$ (Rosado {\etal}1981).
Since the object is density-bounded and leaking ionizing
photons (Table~\ref{tabsample}), the shell is probably fully
ionized, which helps explain the lack of corresponding \hi\
structure.  However, this would still imply that the \hi\ distribution
is clumpy or patchy, to explain the absence of an \hi\ hole.
Typically in this region \nhi$\sim 1.3\times 10^{21}\cm2$, suggesting a mean
volume density of 0.7 $\cc$.

\section{20-cm continuum observations}

Our 20-cm (1380 MHz) continuum observations of the three superbubbles show
excellent spatial correspondence with the \Ha\ distributions, as
expected for the nebular thermal continuum.  We measure flux densities
within our 128 MHz bandwidth of $84\pm 17$ mJy and $173\pm 35$ mJy for
DEM L25 and DEM L301, respectively, excluding point sources.  The flux
density of DEM L50, excluding one point source, is $229\pm 46$ mJy,
of which $99\pm 20$ mJy corresponds to the SNR in the northern region.

Since all three superbubbles have been suggested to be affected by
shocks, there may be a non-negligible contribution to the continuum
from non-thermal emission.  To evaluate this possibility, we
constructed spectral index maps using our 1380 MHz data and 843 MHz
data from the Molonglo Observatory Synthesis Telescope (MOST) survey 
of the Magellanic Clouds (Turtle {\etal}1998).  The results are noisy,
but broadly suggest that DEM L25 and the SNR N186~D in DEM L50 have
spectral indices roughly around $\alpha \sim -0.8$ for flux density $S
\propto \nu^\alpha$, whereas DEM L301 and the remainder of DEM 
L50 have lower spectral indices $\alpha \sim -0.3$.  This is consistent with
the stronger shock signatures seen for DEM L25 and SNR N186~D,
specifically, the higher \oiii/\Ha, \sii/\Ha, overproduction of \Ha,
and higher-velocity kinematics.  The data thus qualitatively
suggest the presence of some non-thermal emission for at least DEM L25
and SNR N186~D.  These results are consistent with the spectral
indices reported by Filipovi\'c {\etal}(1998) in their multi-band
radio continuum survey.  For DEM L25 and DEM L50 (unresolved), they
obtain $\alpha = -0.67 \pm 0.09$ and $\alpha = -0.73 \pm 0.19$,
respectively, while for DEM L301 they find $\alpha = -0.02 \pm 0.03$.
These values may be compared to the mean $\alpha = -0.15$, standard
deviation $\sigma= 0.31$ for \hii\ regions; and mean $\alpha = -0.43$,
$\sigma= 0.19$ for SNRs, reported by the Filipovi\'c {\etal}(1998)
survey.

\section{Conclusion}

We have obtained \hi\ observations of the environment around three
optically-selected superbubbles in the LMC.  The optical nebulae have
similar properties:  all are roughly 50 pc in radius and show
optical kinematic structures with anomalously high expansion
velocities that contrast with values expected from the standard
evolutionary model for wind and SN-generated superbubbles.  We find
that the \hi\ 
environment within roughly 500 pc of the objects vary to an extreme.
DEM L25 shows no \hi\ associated with the superbubble itself, but
appears to be nestled within an \hi\ hole, and the shell is interacting 
with an adjacent \hi\ cloud; DEM L50 shows an \hi\
component to the shell that is $\sim 10$ times more massive than the
ionized shell, but is otherwise in a large \hi\ void; and DEM
L301 shows no correlation of any kind between the optical and \hi\
distributions.  Furthermore, the high-velocity
kinematics observed optically for all three objects are not seen in
any of our \hi\ data.  These findings strongly caution against inferring
properties of the neutral interstellar environment around any individual
superbubbles without direct observations. 

All three superbubbles have reported signatures suggestive of shock
activity, including supersonic nebular velocities.
DEM L25 and DEM L50 show elevated \Ha\ emission in excess of that 
predicted from photoionization by their classified stellar
populations, and DEM L50 encompasses the known SNR N186~D on its
northern edge.  DEM L25 also shows elevated \oiii/\Ha\ and all objects
show high \sii/\Ha.  The emission-line spectrum of DEM L301 was
modeled in detail by Oey {\etal}(2001), who found evidence of
contributions by shock excitation.  We examined our 1380 MHz (20 cm)
continuum in conjunction with 843 MHz continuum observations from the
MOST survey to obtain rough, spatially-resolved estimates of the
spectral indices.  These suggest some presence of a non-thermal
component for at least DEM L25 and the SNR N186~D.

We mentioned above that the new \hi\ data do not show the complex,
high-velocity shell components that are evident in the optical nebular
data for these objects.  However, the uneven \hi\ distributions for DEM
L25 and DEM L301 are consistent with the high ionization state of these
nebulae, suggesting that these shells are fully ionized.  Their patchy
\hi\ remains consistent with the optical filaments corresponding to
incipient blowout features or material accelerated by SNR impacts, as
has been previously suggested.  The lack of correspondence between the
nebular and \hi\ kinematics is more problematic for DEM L50, although
the dominant \Ha\ and \hi\ components are broadly consistent with each
other.  The \hi\ velocity profile suggests that the neutral component
may correspond to the approaching side of the shell.

\acknowledgments
We gratefully acknowledge Anne Green for providing the MOST
survey data, and Sung Kim for assistance during one of our observing
runs.  MSO thanks the ATNF/CSIRO for warm hospitality through their
Distinguished Visitor Programme.  Some of this work was 
carried out by MSO while at the Institute of Astronomy, Cambridge and
the Space Telescope Science Institute.  We also thank the referee, 
Margarita Rosado, for her helpful suggestions.

\clearpage

\begin{figure}
\caption{Composite narrow-band images of $a$) DEM L25 and DEM L50; and
$b$) DEM L301.  \Ha\ is shown in green, \oiii\ in blue, and \sii\ in
red.  Figure~\ref{figopt} is constructed from the
continuum-subtracted images with arbitrary intensity scaling. 
Beware that the relative intensities of the three bands vary between
the two fields.  North is up, east to the left; see
Figures~\ref{fignhi}, \ref{figd25}, \ref{figd50},
and \ref{figd301} for angular scale.
\label{figopt}}
\end{figure}

\begin{figure}
\caption{\hi\ column density maps of the environment around $a$) DEM
L25 and DEM L50; and 
$b$) DEM L301.  The circles roughly indicate the position and
size of the objects, with DEM L25 on the west side of Figure~\ref{fignhi}$a$.
\label{fignhi}}
\end{figure}

\begin{figure}
\caption{Composite images of $a$) DEM L25 and $b$) DEM L50, with \Ha\
shown in red, \oiii\ in green, and \hi\ in blue, using arbitrary
intensity scaling.  
\label{figopthi}}
\end{figure}

\begin{figure}
\caption{\hi\ maps of the region around DEM L25 and L50.  Each panel
shows the average of four channels, whose mean heliocentric velocity
in $\kms$ is shown in the upper left.  The circles indicate the
positions of DEM L25 and L50, with the former on the west side.
\label{figd25chan}}
\end{figure}

\begin{figure}
\caption{\hi\ maps for the region around DEM L301.  Each panel shows
the average of five velocity channels.
\label{figd301chan}}
\end{figure}

\begin{figure}
\caption{\nhi\ contour map of DEM L25, superimposed on \Ha.  Contour
intervals are $2\times10^{20}\ \cm2$, beginning at $2\times10^{19}\ \cm2$.
\label{figd25}}
\end{figure}

\begin{figure}
\caption{Position-velocity diagram oriented E-W across the cloud to
the W of DEM L25.  The center is at 04 53 14.7, --69 59 35, with E to
the left (negative offsets).
\label{figd25pvcloud}}
\end{figure}

\begin{figure}
\caption{Position-velocity diagram for the \hi\ mass NE of DEM L25,
centered at 04 54 29.9, --69 55 35 (J2000).  N is to the left
(negative offsets).
\label{figd25pvne}}
\end{figure}

\begin{figure}
\caption{\nhi\ contour map of DEM L50, superimposed on \Ha.  Contours
are as in Figure~\ref{figd25}.
\label{figd50}}
\end{figure}

\begin{figure}
\caption{Position-velocity diagrams across the center of DEM L50,
oriented $a$) E-W; and $b$) N-S.  E and N are to the left (negative offsets).
\label{figd50pv}}
\end{figure}

\begin{figure}
\caption{Position-velocity diagram across the SNR N186~D.  The slice
is oriented N-S, centered at 05 00 12.3, --70 06 23 (J2000), with N
shown to the left.
\label{figd50pvsnr}}
\end{figure}

\begin{figure}
\caption{\nhi\ contour map of DEM L301, superimposed on \Ha.  Contour
intervals are $3\times10^{20}\ \cm2$, beginning at $2\times10^{19}\ \cm2$.
\label{figd301}}
\end{figure}

\begin{figure}
\caption{Position-velocity diagram across the center of DEM L301,
oriented $a$) E-W; and $b$) N-S.  E and N are shown to the left.
\label{figd301pv}}
\end{figure}

\vfill\eject
\pagebreak
\clearpage

%
%
%
%
%
%
%
%
%
%
%

\setcounter{figure}{6}

\begin{figure*}
\epsscale{0.5}
\plotone{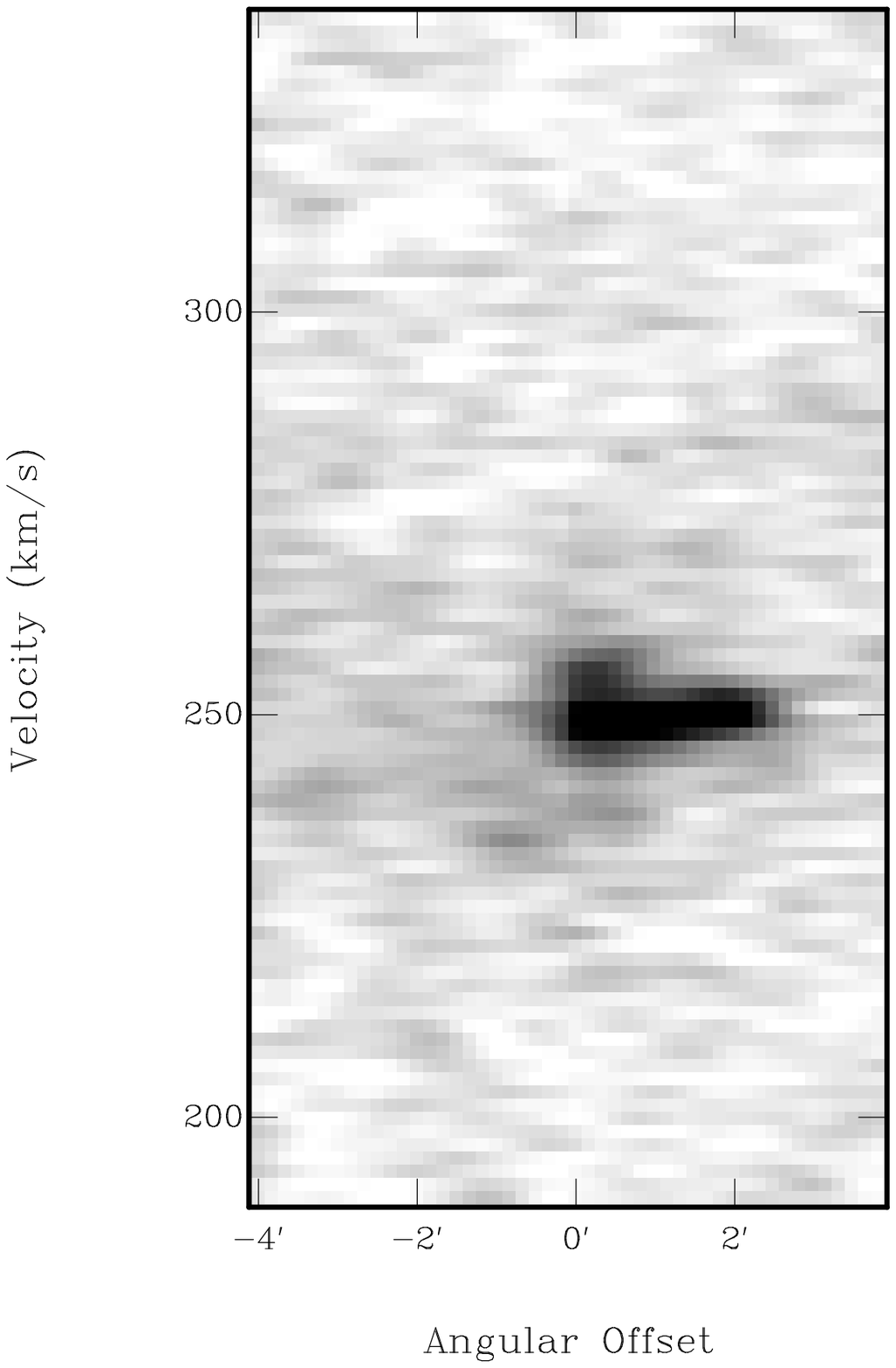}
\caption{}
\end{figure*}

\begin{figure*}
\epsscale{0.5}
\plotone{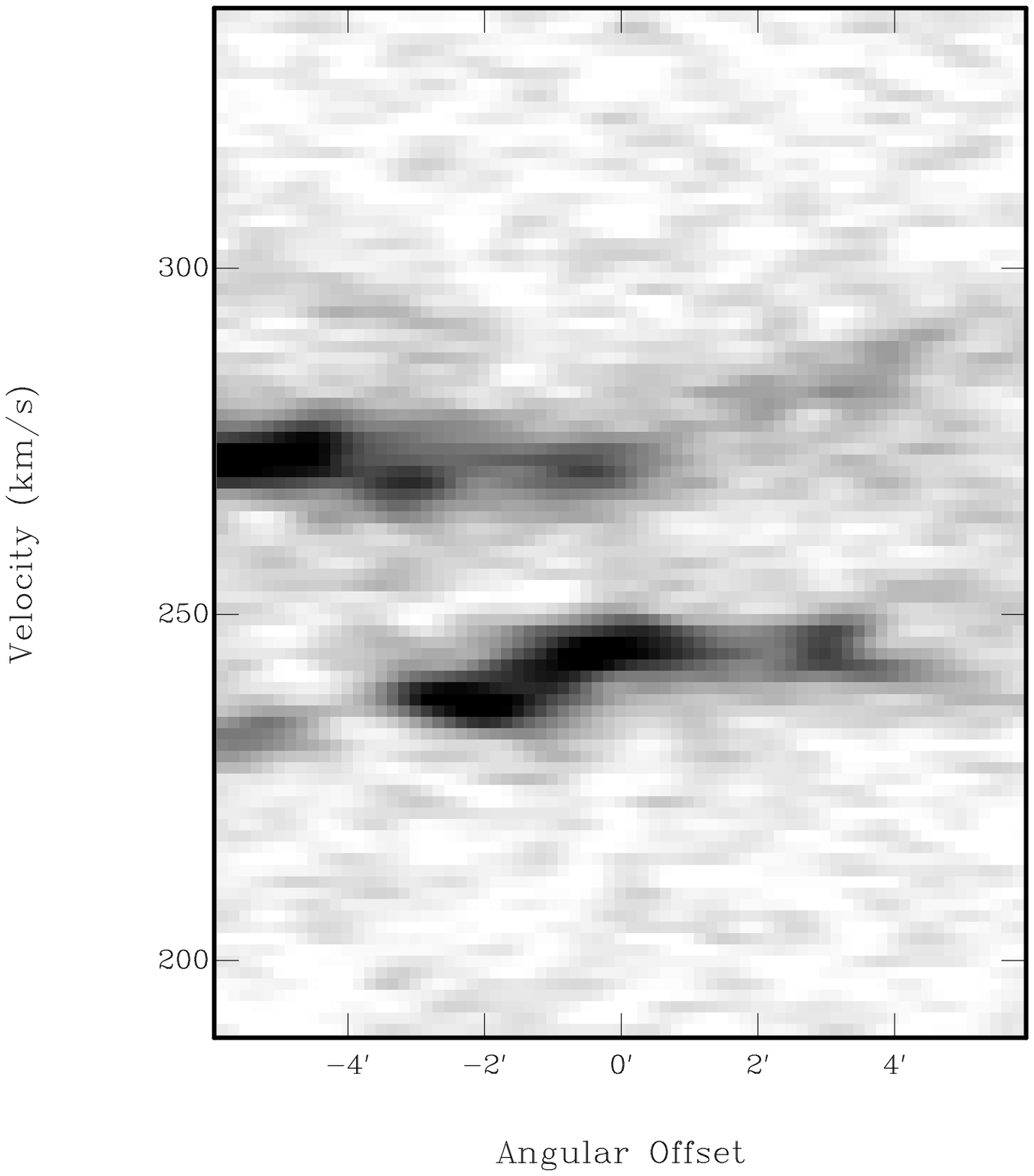}
\caption{}
\end{figure*}


\setcounter{figure}{9}

\begin{figure*}
\epsscale{0.5}
\plotone{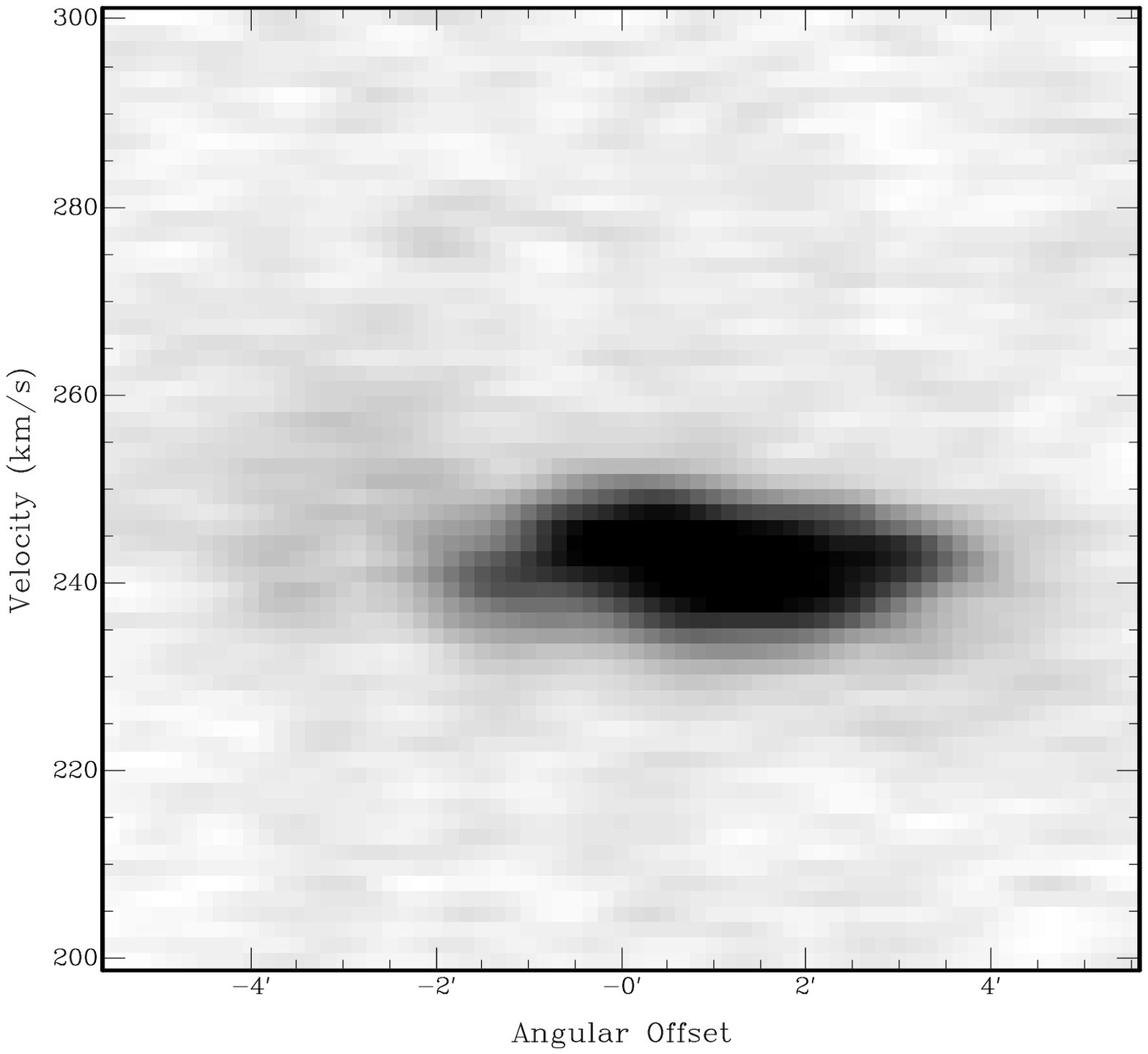}
\plotone{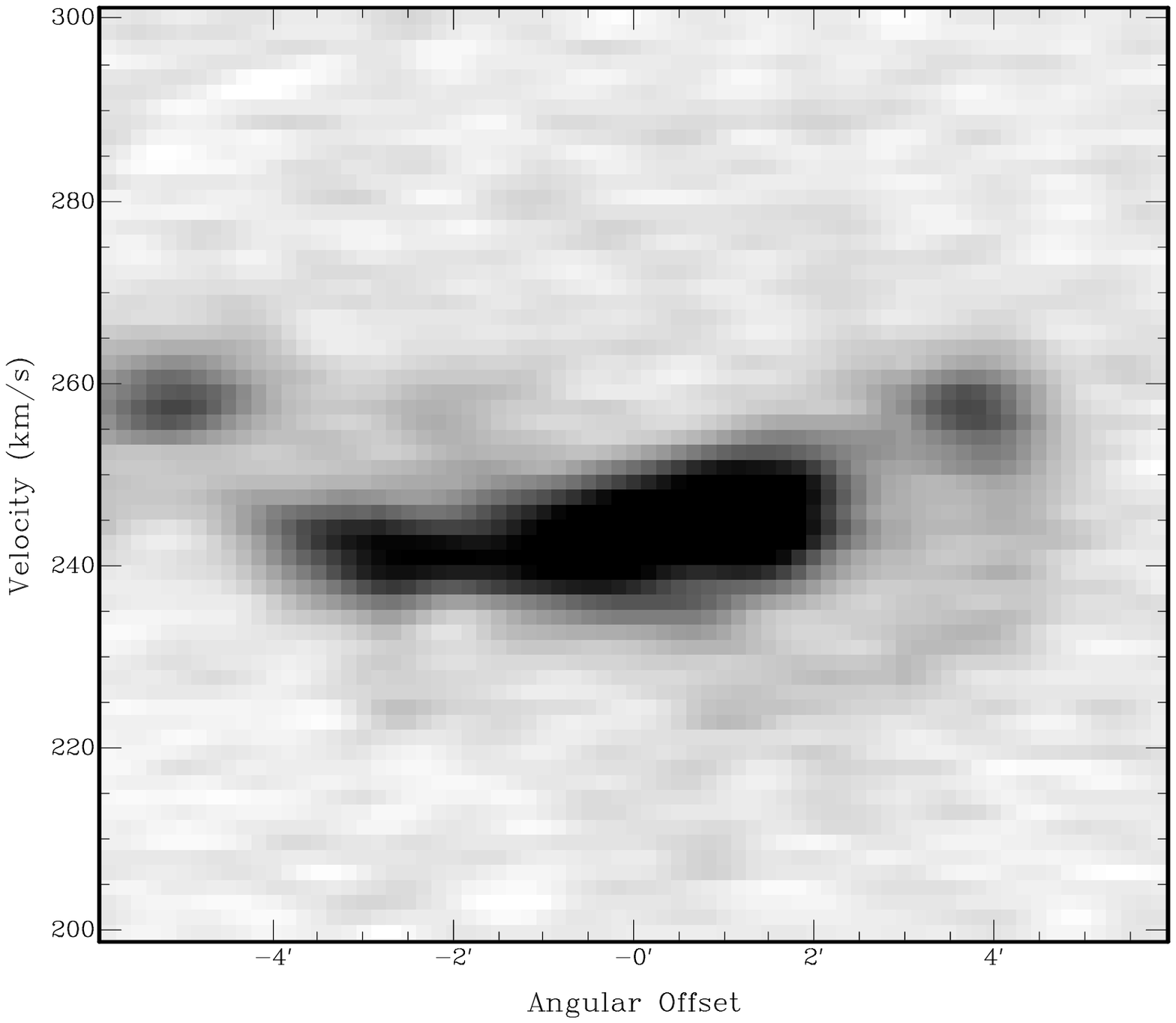}
\caption{$a$ and $b$}
\end{figure*}

\begin{figure*}
\epsscale{0.5}
\plotone{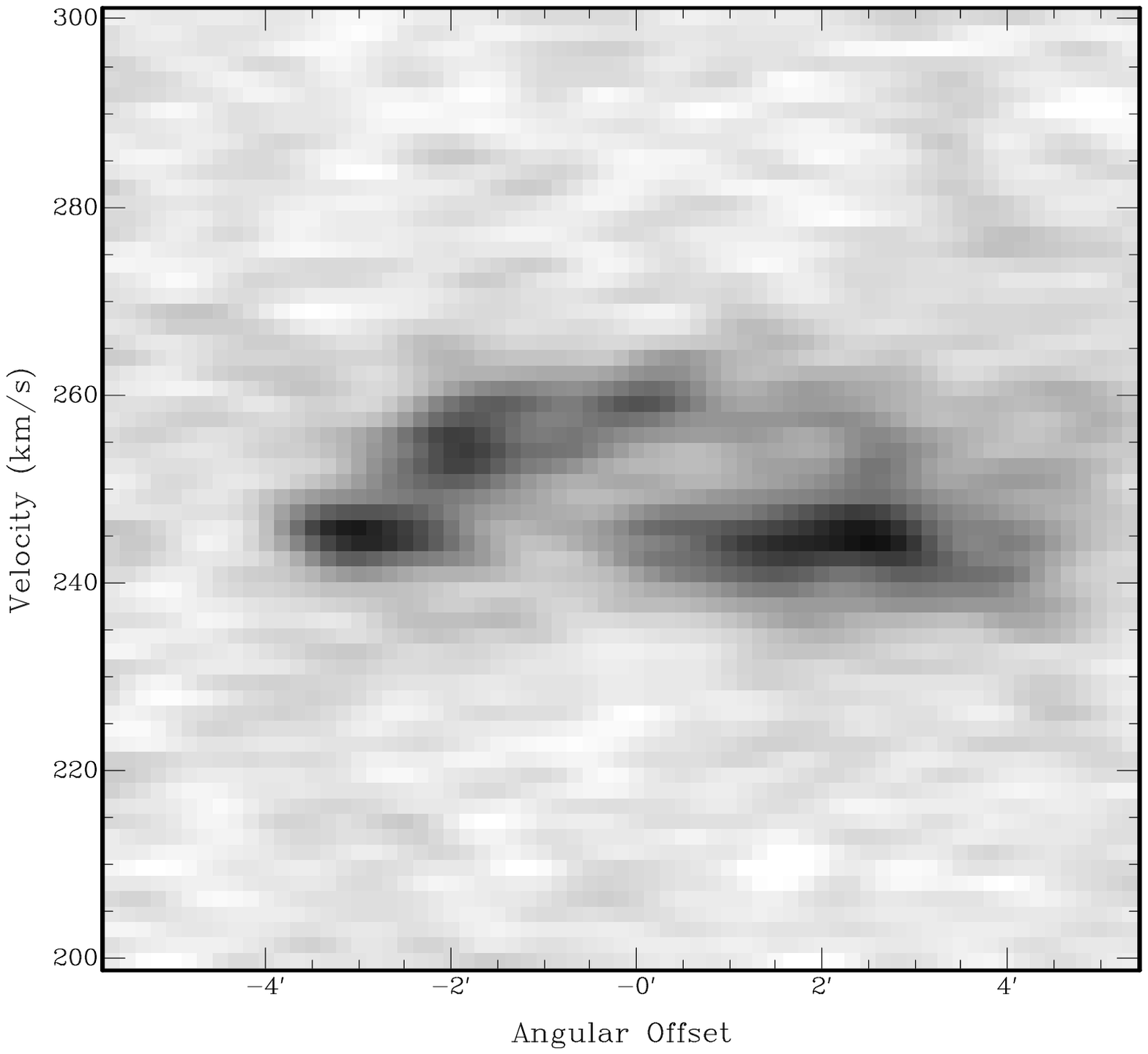}
\caption{}
\end{figure*}


\setcounter{figure}{12}

\begin{figure*}
\epsscale{0.5}
\plotone{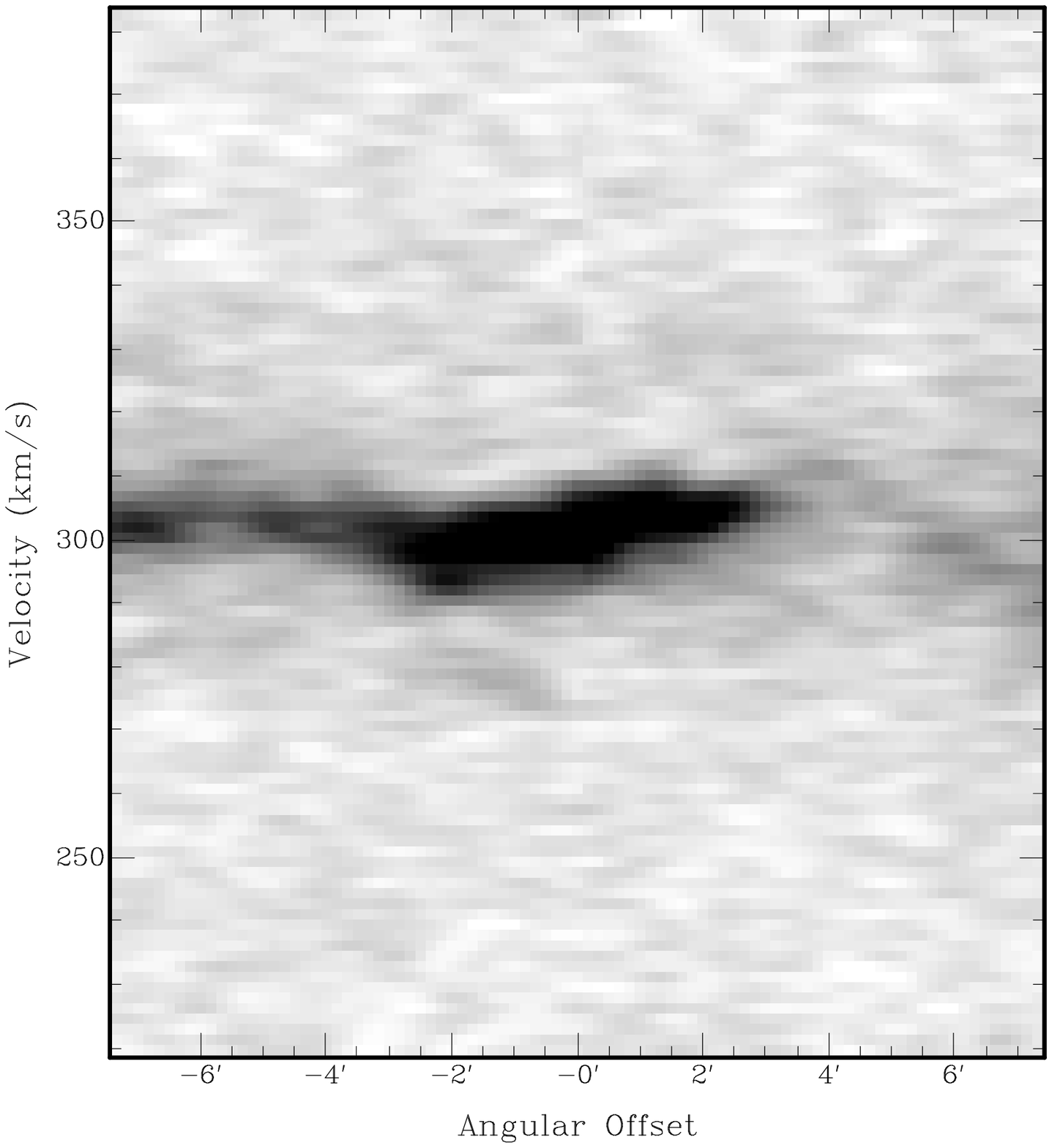}
\plotone{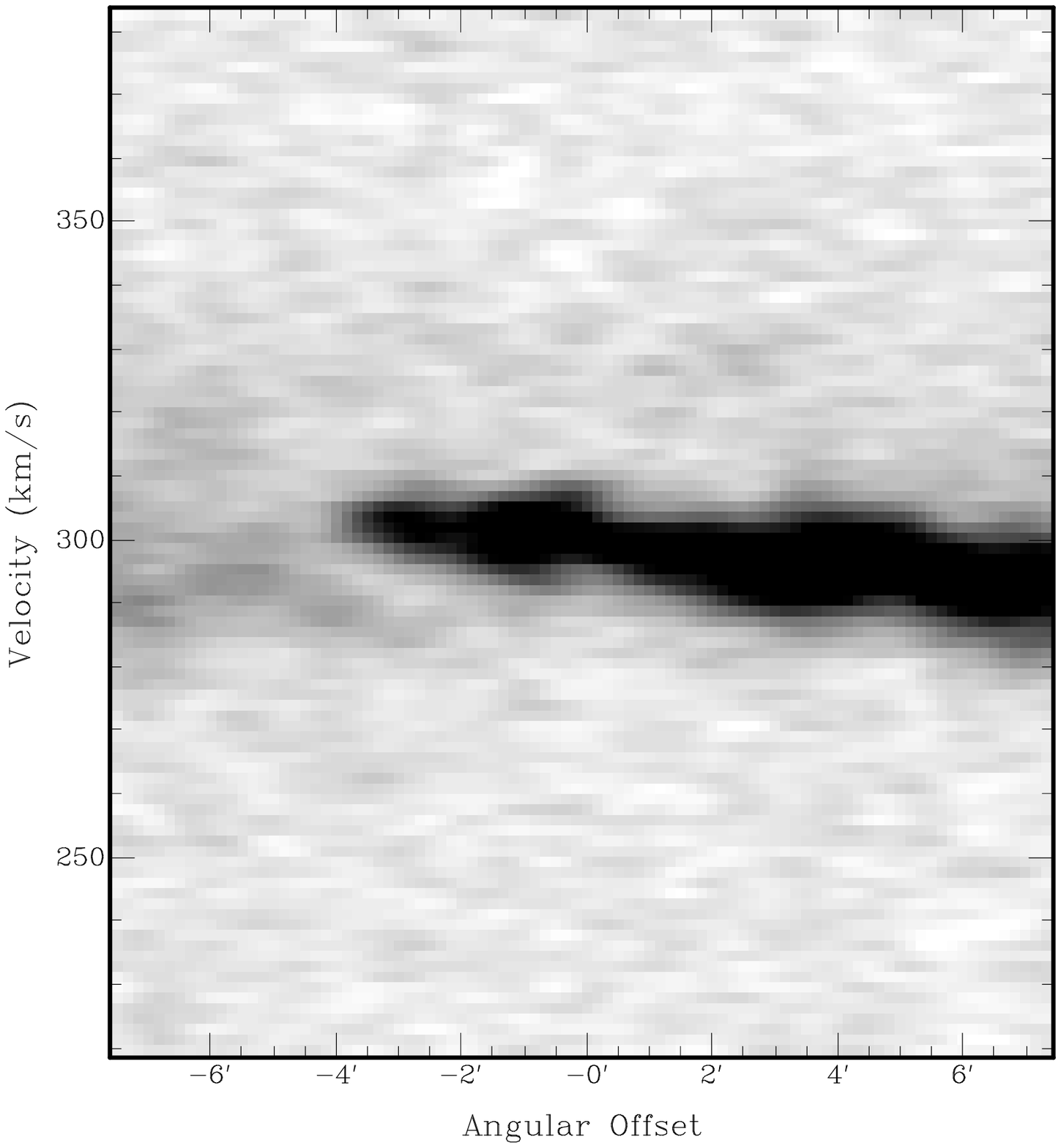}
\caption{$a$ and $b$}
\end{figure*}

\end{document}